\documentclass[preprint,12pt]{elsarticle}
\usepackage{appendix}
\usepackage{longtable}
\usepackage{epsfig} 
\usepackage{multicol}
\usepackage{amsmath}
\usepackage{epsfig} 
\usepackage{multicol}
\usepackage{nomencl}

\usepackage{amsmath}
\usepackage[flushleft]{threeparttable}
\usepackage{adjustbox}
\usepackage{nomencl}
\usepackage{etoolbox}
\usepackage{framed} 

\usepackage{lineno}

\begin{document}
\begin{frontmatter}
\title{A three-dimensional thermal model of the human cochlea for magnetic cochlear implant surgery}

\author{Fateme Esmailie}\author{Mathieu Francoeur}\author{Tim Ameel\corref{mycorrespondingauthor}}
\address{Department of Mechanical Engineering, University of Utah, Salt Lake City, Utah 84112, USA}
\cortext[mycorrespondingauthor]{Corresponding author}
\ead{ameel@mech.utah.edu}


\begin{abstract}
In traditional cochlear implant surgery, physical trauma may occur during electrode array insertion.  Magnetic guidance of the electrode array has been proposed to mitigate this medical complication. After insertion, the guiding magnet attached to the tip of the electrode array must be detached via a heating process and removed. This heating process may, however, cause thermal trauma within the cochlea. In this study, a validated three-dimensional finite element heat transfer model of the human cochlea is applied to perform an intracochlear thermal analysis necessary to ensure the safety of the magnet removal phase. Specifically, the maximum safe input power density to detach the magnet is determined as a function of the boundary conditions, heating duration, cochlea size, implant electrode array radius and insertion depth, magnet size, and cochlear fluid. A dimensional analysis and numerical simulations reveal that the maximum safe input power density increases with increasing cochlea size and the radius of the electrode array, whereas it decreases with increasing electrode array insertion depth and magnet size. The best cochlear fluids from the thermal perspective are perilymph and a soap solution. Even for the worst case scenario in which the cochlear walls are assumed to be adiabatic except at the round window, the maximum safe input power density is larger than that required to melt 1 $\rm{mm^3}$ of paraffin bonding the magnet to the implant electrode array. By combining the outcome of this work with other aspects of the design of the magnetic insertion process, namely the magnetic guidance procedure and medical requirements, it will be possible to implement a thermally safe patient-specific surgical procedure.    
\end{abstract}

\end{frontmatter}

\makenomenclature
\nomenclature[A, 01]{$A$}{Surface area $\left[\rm{mm^2}\right]$}
\nomenclature[A, 02]{$a$}{Constant coefficient for input power density equation $\left[-\right]$}
\nomenclature[A, 03]{$B$}{Thermal dose constant $\left[-\right]$}
\nomenclature[A, 04]{CEM}{Cumulative Equivalent Minutes $\left[\rm{s}\right]$}
\nomenclature[A, 05]{$c_p$}{Specific heat $\left[\frac{\rm{J}}{\rm{kg} \cdot \rm{K}}\right]$}
\nomenclature[A, 06]{$D$}{Magnet diameter $\left[\rm{mm}\right]$}
\nomenclature[A, 07]{$k$}{Thermal conductivity $\left[\frac{\rm{W}}{\rm{m \cdot K}}\right]$}
\nomenclature[A, 08]{$L$}{Magnet length $\left[\rm{mm}\right]$}
\nomenclature[A, 09]{$L_c$}{Characteristic length $\left[\rm{mm}\right]$}
\nomenclature[A, 10]{$N$}{Total numbers of time intervals $\left[-\right]$}
\nomenclature[A, 12]{$q$}{Power density $\left[\frac{\rm{W}}{\rm{m^3}}\right]$}
\nomenclature[A, 13]{$\mathbf{{q_r^{\prime\prime}}}$}{Radiative flux $\left[\frac{\rm{W}}{\rm{m^2}}\right]$}
\nomenclature[A, 14]{$\dot{Q}$}{Heat transfer rate $\left[{\rm{W}}\right]$}
\nomenclature[A, 15]{$r$}{Electrode array radius $\left[\rm{mm}\right]$}
\nomenclature[A, 17]{$t$}{Time $\left[\rm{s}\right]$}
\nomenclature[A, 18]{$t_{n}$}{Length of $n^{th}$ time interval $\left[\rm{s}\right]$}
\nomenclature[A, 19]{$T$}{Temperature $\left[\rm{K}\right]$}
\nomenclature[A, 20]{$\mathbf{u}$}{Velocity $\left[\frac{\rm{m}}{\rm{s}}\right]$}
\nomenclature[A, 21]{$\dot{U}$}{Internal energy rate $\left[\rm{W}\right]$}
\nomenclature[A, 22]{$V$}{Volume of cochlea $\left[\rm{mm^3}\right]$}
\nomenclature[A, 23]{\textbf{Greek symbols}}{}
\nomenclature[B, 01]{$\alpha$}{Thermal diffusivity $\left[\rm{\frac{m^2}{s}}\right]$}
\nomenclature[B, 02]{$\rho$}{Density $\left[\frac{\rm{kg}}{\rm{m^3}}\right]$}
\nomenclature[B, 03]{$\omega$}{Blood perfusion rate $\left[\frac{\rm{1}}{\rm{s}}\right]$}
\nomenclature[C, 01]{\textbf{Subscripts}}{}
\nomenclature[C, 02]{$\rm{bl}$}{Blood}
\nomenclature[C, 03]{$magnet$}{Magnet}
\nomenclature[C, 04]{$met$}{Metabolic}
\nomenclature[C, 05]{$n$}{Number of the time interval}

\printnomenclature

\pagenumbering{arabic}

\section{Introduction}

Magnetic cochlear implant surgery is a promising procedure to reduce potential physical trauma associated with manual insertion of cochlear implants \cite{Clark, Bruns}.~Magnetic guidance of cochlear implants decreases the insertion force by 50$\%$ \cite{Clark, Leon1}. To steer the implant within the cochlea, a magnet is attached to the tip of the implant electrode array (EA) and is guided by an external magnetic field.~After the EA insertion, the magnet may move if the patient is exposed to a strong magnetic field such as in Magnetic Resonance Imaging (MRI)  \cite{majdani,morshed}. In addition, the magnet may cause artifacts on MRI results \cite{majdani,morshed}. To avoid potential medical complications, the magnet must be removed from the cochlea. The removal can be initiated by first detaching the magnet from the implant EA by supplying heat. As such, there is a risk that the magnet detachment process causes thermal trauma within the cochlea. It is therefore critical to analyze heat transfer that occurs in the cochlea during the magnet detachment and removal process. 

Heat transfer within the cochlea has been studied in applications such as infrared neural stimulation \cite{Thompson2013InfraredNS,THOMPSON201546, rajguru4-INS, Liljemalm, xia} and therapeutic hypothermia \cite{Rajgurumain, rajguru2-terapeutic, rajguru3-therapeutic}. As an initial attempt to evaluate the thermal impact of the magnet detachment process on cochlear tissues, a simplified uncoiled model of the cochlea with an inserted implant EA has been developed \cite{uncoiledcochlea}.~The heat transfer analysis of the uncoiled model showed that the implant EA acts as a fin and enhances thermal transport to the surroundings. The results also revealed that the relative contribution of natural convection is negligible owing to the implant EA driving perilymph out of the cochlea. A three-dimensional (3D) COMSOL Multiphysics heat transfer model of the scala tympani was later developed and validated using a phantom subjected to a sudden change in its thermal environment and localized heating \cite{cochleasecondpaper}. The maximum error of the model was found to be less than 6$\%$. In addition, it was confirmed that natural convection has a negligible impact on the thermal management of the cochlea during the magnet detachment process. Despite these efforts, a heat transfer analysis within a 3D model of the entire human cochlea that includes the three canals (scala tympani, scala vestibuli, scala media) is still lacking. 

The objective of this paper is to determine the maximum safe input power density to detach the magnet during cochlear implant surgery in a 3D model of the cochlea as a function of the cochlea boundary conditions, heating duration, cochlea size, implant EA insertion depth and radius, magnet size, and cochlear fluid. The 3D heat transfer model validated in Ref. \cite{cochleasecondpaper} is applied for that purpose and expanded to the full human cochlea using the geometry developed by Gerber et al. \cite{Gerber}. The rest of the paper is organized as follows.~The physical and mathematical models are described in Section \ref{Description of the physical and mathematical models}. The simulation results and parametric studies are presented in Section \ref{Determination of maximum safe input power density to detach the magnet}. Concluding remarks are finally provided.

\section{Description of the physical and mathematical models}
\label{Description of the physical and mathematical models}
\begin{figure}[h]
\begin{center}
  \includegraphics[width=5 in]{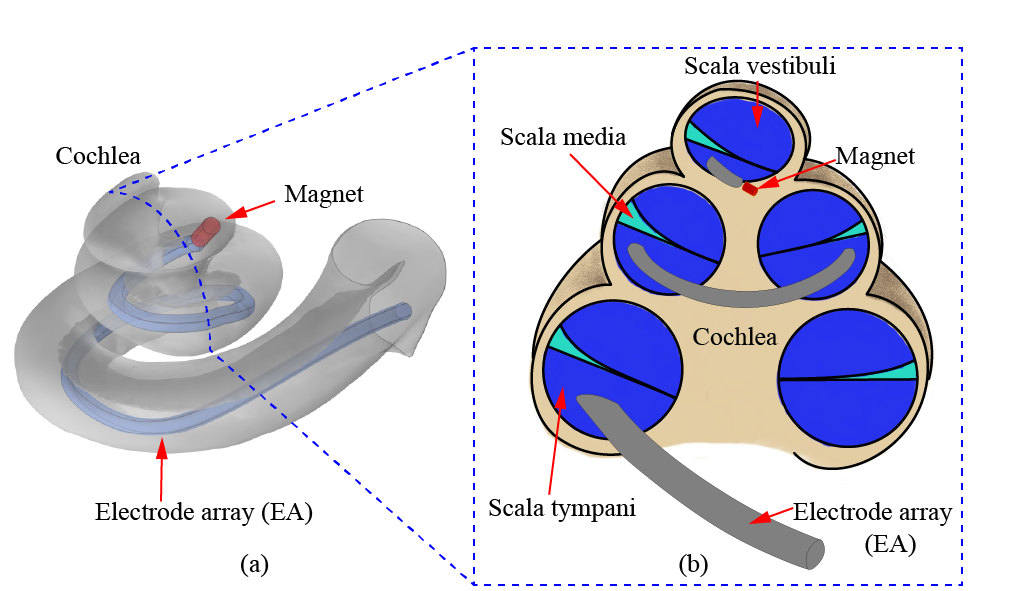}
  \caption{3D model of the cochlea with inserted EA and magnet. (a) 3D physical model of the cochlea provided by Sicas Medical Image Repository \cite{Gerber, Sicas}. The EA and magnet have been added to the 3D cochlea model. (b) Schematic of the cochlea with inserted implant EA and magnet.}
  \label{Cochleawithimplant}
  \end{center}
\end{figure}

    The cochlea consists of three canals: the scala tympani, the scala media, and the scala vestibuli \cite{Yoo, WYSOCKI20011, TOTH2006363,Raufer13977}. The implant EA is inserted in the scala tympani (see Figure~\ref{Cochleawithimplant}) \cite{Yoo, WYSOCKI20011, TOTH2006363,Raufer13977}. The scala tympani and scala vestibuli are filled with perilymph, whereas the scala media is filled with endolymph \cite{Yoo, WYSOCKI20011, TOTH2006363,Raufer13977}. Endolymph and perilymph are dilute saline fluids and their thermophysical properties are assumed to be identical to those of water (see Table \ref{table}) \cite{Rajgurumain, Kassemi:2005}. The 3D physical model of the cochlea used in this work has been developed by Gerber et al.\cite{Gerber} via computerized tomography scans of 52 temporal bones. The model includes all three cochlear canals, but the separating membranes are omitted (i.e., the scala tympani, scala media, and scala vestibuli form a single canal). The model is available on the Sicas Medical Image Repository website \cite{Sicas}. The physical model is converted to IGES format and imported into COMSOL Multiphysics 5.5 for the heat transfer simulations.~The implant EA and magnet are added to the cochlea in COMSOL Multiphysics 5.5. The volume of the cochlea is 87.20 $\rm{mm^3}$. The implant EA is modeled as a spiral object (2.825 $\rm{mm^3}$ volume) with a 0.2-mm-radius circular cross section and a dome-like tip. A 1-mm-long, 0.25-mm-radius cylindrical magnet is located at the tip of the EA. The thermophysical properties of the implant EA and magnet are listed in Table \ref{table}.

\begin{table}[h]
\caption{Thermophysical properties of the substances used in the simulations }
\label{table}
\scriptsize
\begin{center}
\begin{tabular}{l l l l} 
    \multicolumn{4}{c}{}\\
\hline
Substance& 
Density $\rho$~[$\frac{\rm{kg}}{\rm{m^3}}$]& Thermal conductivity $k$ [$\frac{\rm{W}}{\rm{m \cdot K}}$]& Specific heat $c_p$ [$\frac{\rm{J}}{\rm{kg \cdot K}}$]\\
\hline
Perilymph (Water) &992.20 \cite{comsol} &0.625\cite{comsol} &4176.6\cite{comsol} \\
Magnet   & 430$^{\mathrm{a}}$ & 8.1$^{\mathrm{a}}$ & 7500$^{\mathrm{a}}$ \\
Electrode array (EA) & 19400$^{\mathrm{b}}$ & 28$^{\mathrm{b}}$ & 127.7$^{\mathrm{b}}$ \\
Blood & 1050 \cite{Rajgurumain} & 0.52\cite{Rajgurumain}& 3840\cite{Rajgurumain}\\
Bone & 1908 \cite{Rajgurumain} & 0.32\cite{Rajgurumain}& 1313\cite{Rajgurumain}\\
Air  & 1.13 \cite{comsol}  & 0.027 \cite{comsol} & 1006.4 \cite{comsol} \\
Glycerol & 1260.00 \cite{glycerolcp} & 0.285 \cite{glycerol} & 2240.0 \cite{glycerol} \\
Soap solution   & 999 \cite{soap} & 0.60 \cite{soap1} & 4000 \cite{soap}\\
\hline
\multicolumn{3}{p{240pt}}{$^{\mathrm{a}}$Provided by the manufacturer (SUPERMAGNETMAN).

$^{\mathrm{b}}$Calculated based on the information provided by MED-EL \cite{DHANASINGH201793}.
}
\end{tabular}
\label{table}
    
\end{center}
\end{table}
To evaluate the transient temperature change within the cochlea during the magnet detachment process, the energy balance equation is solved using a validated 3D heat transfer model implemented in COMSOL Multiphysics 5.5 \cite{cochleasecondpaper}.~A general form of the energy balance equation in the cochlea is given by \cite{Pennes}:
\begin{equation}
\begin{split}
\rho c_p {\frac{\partial T}{\partial t}}+{\rho c_p \mathbf{u} \cdot \nabla T}+\nabla \cdot \left(-k \nabla T \right)+\nabla \cdot {\mathbf{q}^{\prime\prime}_r}+\rho_{\rm{bl}} c_{p,{\rm{bl}}} \omega_{\rm{bl}} \left(T-T_{\rm{bl}}\right) = q_{\rm{met}}+ q
\end{split}
\label{Pennes'}\end{equation}\noindent
On the left-hand side of Eq. \eqref{Pennes'}, the first term is the thermal energy storage, whereas the subsequent terms are respectively heat transport by convection, conduction, radiation, and blood perfusion. The term $q_{\rm{met}}$ is the volumetric metabolic heat generation, and $q$ is the external volumetric heat source applied to detach the magnet from the implant EA.  

Radiation heat transfer is negligible because of the low temperatures involved (around the body core temperature of 37$^{\circ}$C) within the cochlea during the magnet detachment process \cite{cochleasecondpaper}.~The implant EA forces some of the perilymph out of the cochlea and, consequently, reduces the impact of natural convection to a negligible value with respect to conduction heat transfer \cite{uncoiledcochlea,cochleasecondpaper}. Perfusion and metabolic heat generation are both negligible in comparison to the magnitude of the heat removed by conduction \cite{cochleasecondpaper}. Therefore, the following simplified energy balance equation is solved to simulate heat transfer within the cochlea with an inserted implant EA and magnet:
 \begin{equation} \label{magnetenergy}
{\rho c_p \frac{\partial T}{\partial t}} + {\nabla \cdot (-k \nabla T)} = q
\end{equation}\noindent
The volumetric heat source $q$ is non-zero only in the magnet. 

\begin{figure}[h]
\begin{center}
  \includegraphics[width=4 in]{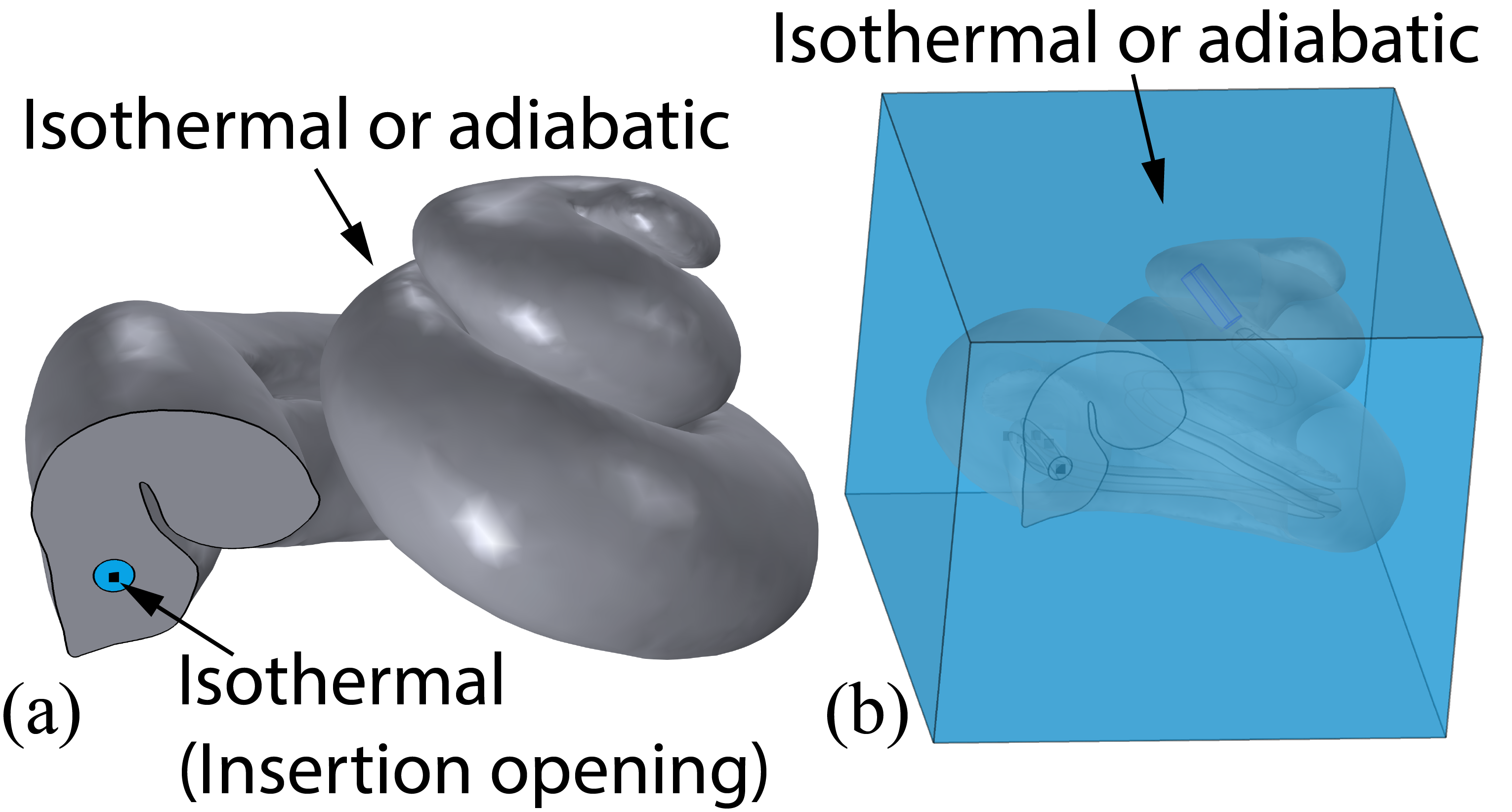}
  \caption{Boundary conditions used in the heat transfer simulations. (a)  The outer boundary of the cochlea may be adiabatic or isothermal, except the insertion opening that it always isothermal at the body core temperature of 37$^{\circ}$C. (b) Cochlea embedded in a temporal bone modeled as a 8 mm $\times$ 8 mm $\times$ 7 mm box with adiabatic or isothermal outer boundary conditions.}
  \label{boundaries}
  \end{center}
\end{figure}

Both isothermal and adiabatic conditions at the cochlea outer boundary are considered in the simulations (see Figure \ref{boundaries}(a)).~Thermoregulatory mechanisms adjust the body core temperature to 37$^{\circ}$C. As such, an isothermal boundary condition is a reasonable assumption. In reality, the cochlea is embedded in a $\sim$ 4-mm-thick \cite{Mahinda2009VariabilityIT,Brutto, kwon} temporal bone characterized by a low thermal conductivity of 0.32 $\rm{\frac{W}{m \cdot K}}$ \cite{Rajgurumain}. This suggests that it is also reasonable to assume that the outer boundary of the temporal bone is adiabatic. The isothermal boundary condition at the bone edges is also physically meaningful due to thermoregulatory effects. Here, the temporal bone is modeled as a 8 mm $\times$ 8 mm $\times$ 7 mm rectangular box surrounding the cochlea (see Figure \ref{boundaries}(b)) with an outer boundary that is either adiabatic or isothermal. In all simulations, the entire cochlea is assumed to be at a constant and uniform temperature of 37$^{\circ}$C, initially.

 \rm{Prior to determining the maximum safe input power density to detach the magnet, the convergence of the numerical simulations is analyzed for grid and time step independence.~For the case in which the cochlea boundary is adiabatic, grid independence is invetigated by increasing the number of elements in four steps from 122,020 to 1,536,679. The temperature at three different locations is shown as a function of the number of elements (Figure \ref{meshconvergence1}(a)) and as a function of the time step (Figure \ref{meshconvergence1}(b)). Refining the mesh by more than one order of magnitude changes the results by less than 3$\%$, based on a 6$^{\circ}$C reference. Therefore, 1,536,679 elements are used in the simulations. Refining the time step by a factor of ten does not change the maximum temperature by more than 1$\%$. A time step of 0.1 s is used for all simulations. For the isothermal boundary condition and the cochlea embedded inside the bone with adiabatic and isothermal bone edges, (1,506,207), (5,903,082), and (3,326,360) elements are used, respectively, to ensure converged results.  

Two thermal boundary conditions are considered for the initial cochlea analysis; 1) an adiabatic condition on the cochlea outer surface, and 2) a cochlea embedded in temporal bone with adiabatic surfaces. The effect of these boundary conditions on the temperature distribution and the maximum input power density required to achieve a set temperature within the cochlea are of interest. Since the cell death rate for many types of tissues increases above 43$^{\circ}$C \cite{thermaldose1,Rhoon, Yoshida10116}, the maximum temperature in the cochlea after 114 s of heating is set to that value. A uniform initial condition of 37$^{\circ}$C is assumed for both cases. The maximum temperature within the cochlea with adiabatic boundaries reaches 43$^{\circ}$C after 114 s by applying a power density of 20 $\rm{\frac{MW}{m^3}}$ to the magnet (Figure \ref{3D temp profile}(a)). For the embedded cochlea, 77 $\rm{\frac{MW}{m^3}}$ in required (Figure \ref{3D temp profile}(b)). 
In both cases, the maximum temperature is located within the magnet, as expected. The temperature decreases with distance from the magnet and towards the isothermal insertion opening (location where the EA is inserted, see Figure \ref{boundaries}). For the cochlea with an adiabatic boundary, heat can only be transferred along the cochlea because the cochlear walls are adiabatic.~On the contrary, for the cochlea embedded in bone, heat is transferred through the cochlear walls to the bone acting as a heat sink, resulting in a higher rate of heat dissipation from the magnet.~As a consequence, greater input power density is required to increase the magnet temperature to 43$^{\circ}$C (see Figure \ref{3D temp profile}(b)).
\begin{figure}
\begin{center}
  \includegraphics[width=\textwidth]{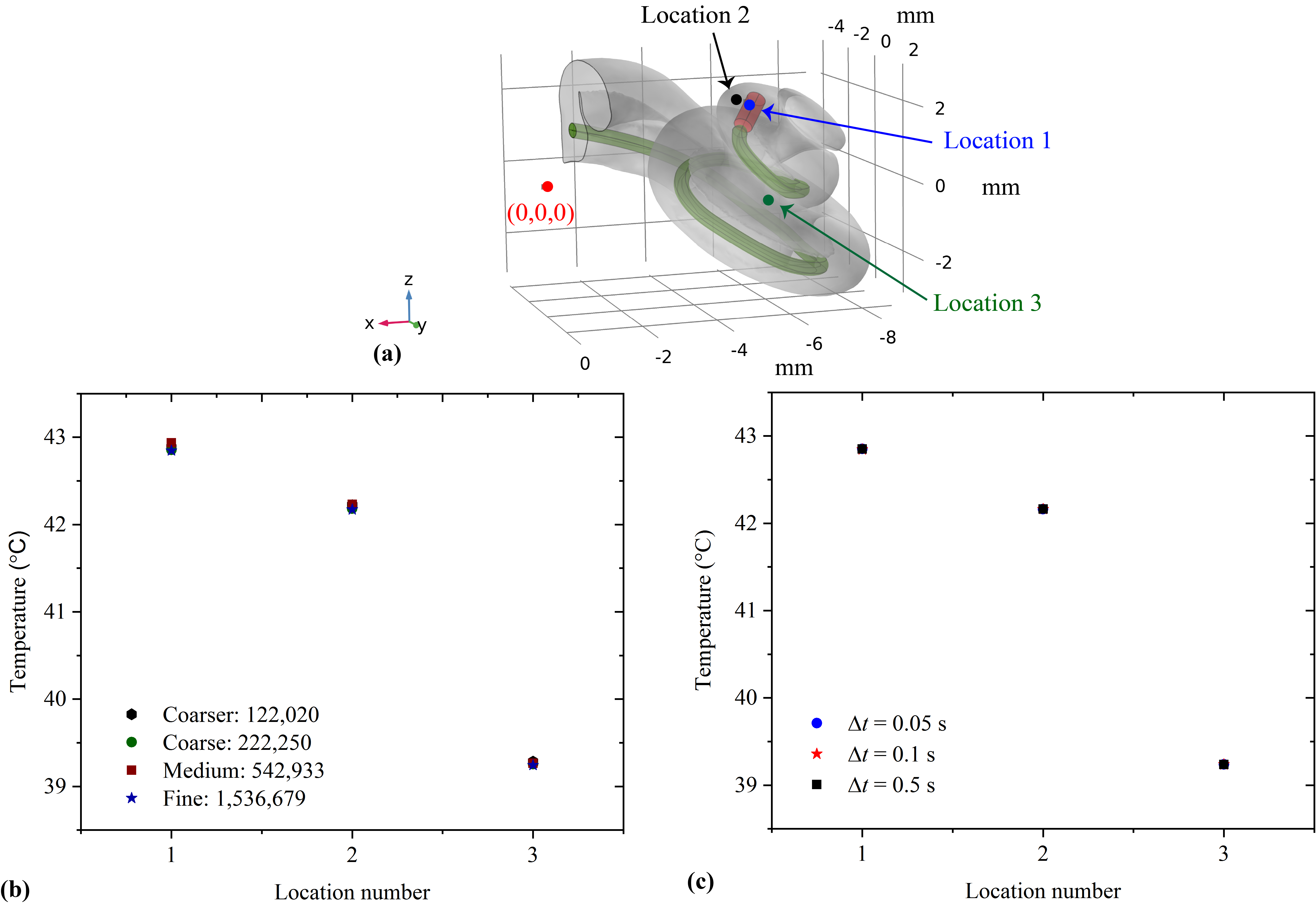}
  \caption{Grid and time independence plots. (a) Location of the sample points. Point 1 corresponds to the location of maximum temperature (-5.5 mm, -0.1 mm, 1.9 mm), while points 2 and 3 are, respectively, located at (-5.2 mm, 0, 2 mm) and (-6 mm, 0, -0.8 mm). In the simulations, the outer boundary of the cochlea is adiabatic, the input power density is $q$ = 20 $\rm{\frac{MW}{m^3}}$, and the data are shown at time $t$ = 114 s. (b) Temperature at three locations in the cochlea as a function of the number of elements. (c) Temperature at three locations in the cochlea as a function of time step.   }
  \label{meshconvergence1}
  \end{center}
\end{figure}

\begin{figure}[h]
\begin{center}
  \includegraphics[width=\textwidth]{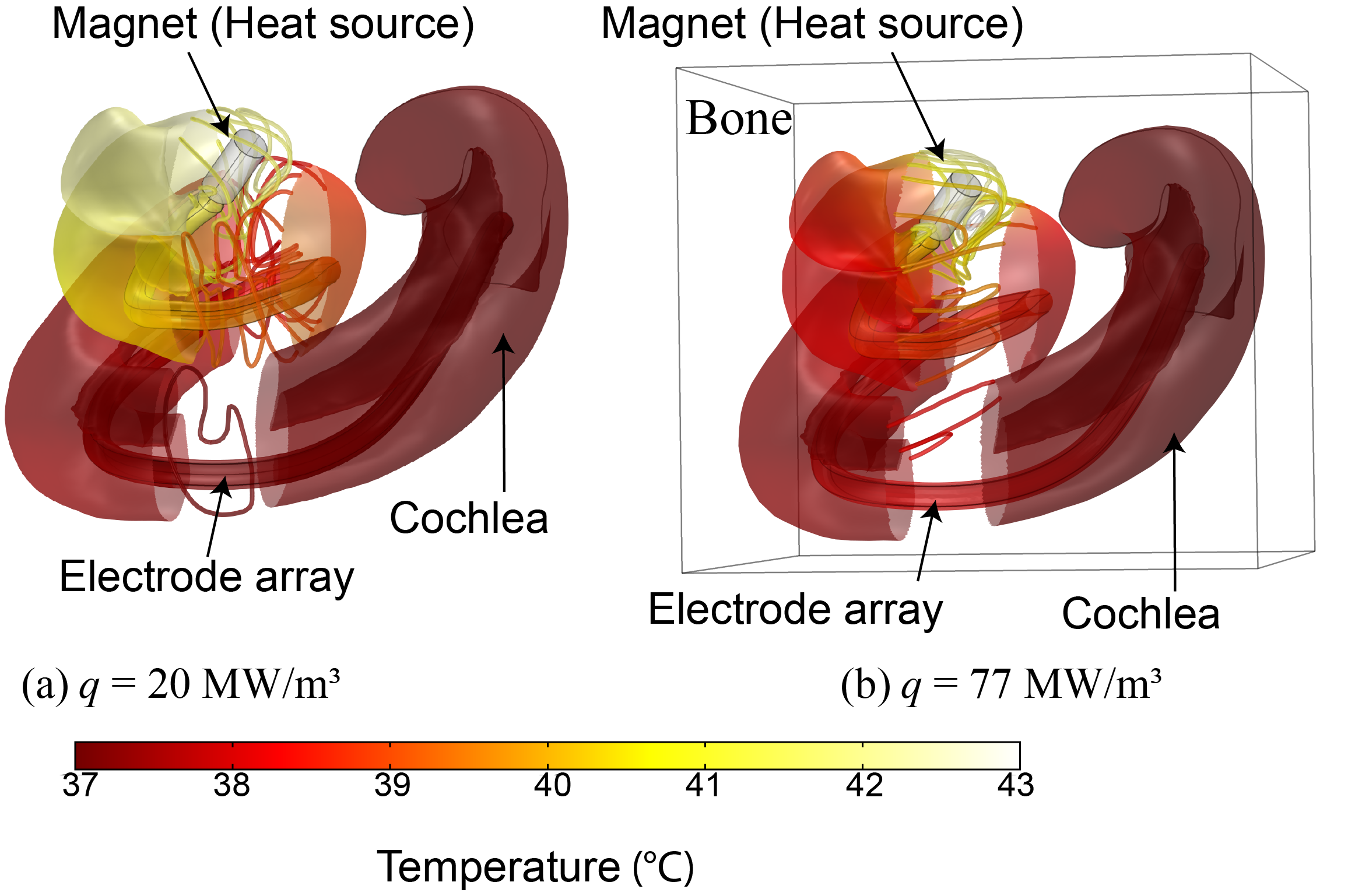}
  \caption{Temperature distribution within the cochlea after 114 s of heating by the magnet. (a) Adiabatic cochlear boundary ($q$ = 20 $\rm{\frac{MW}{m^3}}$). (b) Temporal bone surrounding the cochlea with adiabatic boundary ($q$ = 77 $\rm{\frac{MW}{m^3}}$). Temperature distribution in the middle section is replaced by isotherms for better visualization.}
  \label{3D temp profile}
  \end{center}
\end{figure}

\section{Determination of maximum safe input power density to detach the magnet}
\label{Determination of maximum safe input power density to detach the magnet}
The maximum safe input power density to detach the magnet from the implant EA is a function of cochlea size, the size and insertion depth of the EA, magnet size, thermal boundary conditions, and heating duration.~A potential means to decrease the insertion force, and consequently mitigate the risk of intracochlear physical trauma, is to replace the perilymph by a fluid such as glycerol that decreases the friction inside the cochlea \cite{fluidimpact}. The thermophysical properties of the fluid affects the maximum safe input power density. Therefore, the impact of the cochlear fluid should also be investigated. 

The magnet detachment and removal from the cochlea includes two steps. The first step is to debond and detach the magnet from the EA. The bond may be an adhesive or a shape-memory alloy. In either case, heat must be supplied until  the magnet is released. Thus, once the power to the magnet is terminated and the magnet is released, its temperature is likely to be greater than the body core temperature. During the second step, the hot magnet is removed from the cochlea while the magnet cools.~A moving hot magnet exposes tissues to an elevated temperature for a shorter period of time in comparison to a stationary condition. The worst case scenario, therefore, arises when the hot magnet is motionless while it cools to the body core temperature before it is removed from the cochlea. This worst case scenario (stationary magnet) is considered here to include a safety factor. 

A thermal damage threshold is employed to determine the maximum safe input power density. This thermal threshold is termed the Cumulative Equivalent Minutes (CEM). The CEM at 43$^{\circ}$C (thermal dose) is calculated during magnet heating and cooling to evaluate the maximum safe input power density. According to the literature, $\rm{CEM_{43}}$ for cochlear tissues is 114 s \cite{thermaldose1, Rhoon, Yoshida10116}. This implies that cochlear tissues may be exposed to a temperature of 43$^{\circ}$C for 114 s and then suddenly cooled to the body core temperature without experiencing thermal trauma. In this paper, 114 s is selected as the maximum time the heat source is powered. In practice, the temperature increases and decreases gradually, such that the thermal dose is calculated as follows:
\begin{equation} \label{thermaldose}
\rm{{CEM}_{43}}=\sum_{\it{n}=\rm{1}}^{\it{N}} \it{t}_n \it{B}^{{(\rm{43}-\it{T}_{n})}}
\end{equation}
\noindent
where $t_{n}$ is the time interval that a tissue is exposed to a temperature above 37$^{\circ}$C, and $T_n$ is the average temperature of the $n^{th}$ time interval. $B$ is a constant equal to 0.25 for $T <$ 43$^{\circ}$C and 0.5 for $T >$ 43$^{\circ}$C.
$\rm{CEM_{43}}$ is applied as the thermal threshold to evaluate the maximum safe input power density to detach the magnet in the following sections.

\subsection{Dimensional analysis}
\label{Analytical and numerical parametric study-ch6}
A dimensional analysis is performed to determine analytically the general relationship between maximum input power density and the EA radius and length, cochlea volume, magnet volume, and heating duration. The relationships derived via the dimensional analysis will help interpret the simulation results in Section 3.2. For an adiabatic cochlea, heat released from the magnet is transferred both to the perilymph and the EA. Any heat loss to the surroundings from the round window occurs by conduction in the perilymph and EA. The magnet heat increases the temperature of the magnet itself, the perilymph, and the EA. Energy balances that represent these effects for the magnet, perilymph, and EA are presented in Eqs. (\ref{parametricstudyenergybalancemagnet-ch6}) to (\ref{parametricstudyenergybalanceEA-ch6}). For the dimensional analysis, it is assumed that the thermophysical properties of the perilymph, magnet, and EA are independent of temperature.
\begin{equation}
    \Delta \dot{U}_{magnet} = \dot{Q}_{generation}-\dot{Q}_{perilymph}
    \label{parametricstudyenergybalancemagnet-ch6}
\end{equation}
\begin{equation}
    \Delta \dot{U}_{perilymph} = \dot{Q}_{perilymph}-\dot{Q}_{EA}-\dot{Q}_{perilymph-ambient}
    \label{parametricstudyenergybalanceperilymph-ch6}
\end{equation}
\begin{equation}
    \Delta \dot{U}_{EA} = \dot{Q}_{EA}-\dot{Q}_{EA-ambient}
    \label{parametricstudyenergybalanceEA-ch6}
\end{equation}

$\dot{Q}_{generation}$ is determined by substituting Eqs. (\ref{parametricstudyenergybalanceperilymph-ch6}) and (\ref{parametricstudyenergybalanceEA-ch6}) into Eq. (\ref{parametricstudyenergybalancemagnet-ch6}):
\begin{equation}
    \dot{Q}_{generation} = \Delta \dot{U}_{magnet} +\Delta \dot{U}_{perilymph}+\Delta \dot{U}_{EA}+\dot{Q}_{perilymph-ambient}+\dot{Q}_{EA-ambient}
    \label{Qgeneration}
\end{equation}
The input power density is expressed as $\dot{Q}_{generation} = q V_{magnet}$, where $V_{magnet}$ is the magnet volume. Approximating the time derivative with a ratio of finite differences, the time rates of change of internal energy for the magnet, perilymph, and EA are expressed as $\Delta \dot{U}_{magnet} = \frac{(\rho c_p V \Delta T)_{magnet}}{\Delta t}$, $\Delta \dot{U}_{perilymph} = \frac{(\rho c_p V \Delta T)_{perilymph}}{\Delta t}$, and $\Delta \dot{U}_{EA} = \frac{(\rho c_p V \Delta T)_{EA}}{\Delta t}$, respectively. The heat transferred to the ambient is assumed to be equivalent to the conduction through the ends of the perilymph and EA.  These  terms are approximated as $\dot{Q}_{perilymph-ambient} = (k A \frac{\Delta T}{L})_{perilymph}$ and $\dot{Q}_{EA-ambient}=(k A \frac{\Delta T}{L})_{EA}$. The perilymph volume is determined by subtracting the volumes of the magnet and EA from the cochlea volume $V_{perilymph}= V_{cochlea}-V_{magnet}-V_{EA}$, where the EA is assumed to be a cylinder and the volume of the EA cap (semi-spherical shape) is negligible in comparison to its total volume. The insertion opening is assumed to be circular with radius $r_0$. The maximum safe input power density is now expressed in terms of system variables by substituting the aforementioned expressions into Eq. (\ref{Qgeneration}) and rearranging the equation as follows:
\begin{equation*}
 q= \left[\frac{(\rho c_p \Delta T)_{magnet}}{\Delta t}\right]+\left[\frac{(\rho c_p \Delta T)_{perilymph} (V_{cochlea}-V_{magnet}-(\pi r_{EA}^2 L_{EA}))}{V_{magnet} \Delta t }\right]
 \end{equation*}\\
 \begin{equation*}
 +\displaystyle\left[\frac{(\rho c_p  \Delta T)_{EA} \pi r_{EA}^2 L_{EA}} {V_{magnet} \Delta t}\right] +\left[\frac{(k \Delta T)_{perilymph}\pi (r_0^2-r_{EA}^2)}{V_{magnet} L_{cochlea}}\right]+
  \end{equation*}\\
 \begin{equation}
 \displaystyle\left[\frac{(k \Delta T)_{EA} \pi r_{EA}^2 } {V_{magnet} L_{EA}}\right]\hfill
\label{maximumsafeinputpowerdensity-ch6}
\end{equation}\\

The relationship between the input power density and EA radius, EA length, cochlea volume, magnet volume, and heating duration can be determined by examining  Eq. (\ref{maximumsafeinputpowerdensity-ch6}). 

Assuming temperature-independent thermophysical properties and constant temperature difference, the input power density equation is simplified to:
\begin{equation*}
    q=\left[\frac{a_0-a_1}{\Delta t}\right]+\left[\frac{a_1 V_{cochlea}}{V_{magnet} \Delta t}\right]+\left[\frac{(a_2-a_3) r_{EA}^2 L_{EA}}{V_{magnet} \Delta t}\right]-\left[\frac{a_4 r_{EA}^2}{V_{magnet} L_{cochlea}}\right]+ 
\end{equation*}
\begin{equation}
    \displaystyle\left[\frac{a_5 r_{EA}^2}{V_{magnet} L_{EA}}\right]+ \left[\frac{a_6}{V_{magnet} L_{cochlea}}\right]\hfill
   \label{relations-ch6}
\end{equation}\\

\noindent where: \\
$a_0 = (\rho c_p \Delta T)_{magnet}$, 
$a_1 = (\rho c_p \Delta T)_{perlymph}$,
$a_2 = \pi (\rho c_p \Delta T)_{EA}$,\\
$a_3 = \pi (\rho c_p \Delta T)_{perilymph}$,
$a_4 = \pi (k \Delta T)_{perilymph}$, 
$a_5 = \pi (k \Delta T)_{EA}$, and \\
$a_6$ = $\pi r_0^2 (k \Delta T)_{perilymph}$.\\

\noindent
Based on Eq. (\ref{relations-ch6}), the maximum input power density  is expected to be proportional to the second order of EA radius, proportional to cochlea volume, and inversely proportional to EA length, heating duration, and magnet volume. 

\subsection{Impact of boundary conditions and heating duration}
\label{Impact of boundary conditions and length of heating}
The maximum safe input power density is determined by limiting CEM$_{43}$ to the range $113~\rm{s}$ $<$ CEM$_{43}$ $<$ $114~\rm{s}$. CEM$_{43}$ less than 114 s is safe for cochlear tissues \cite{Yoshida10116}. According to Eq. (\ref{thermaldose}), it may also be thermally safe for cochlear tissues to be exposed to temperatures greater than 43$^{\circ}$C for less than 114 s, implying the maximum temperature within the cochlea may exceed 43$^{\circ}$C. The maximum safe input power density to detach the magnet as a function of the heating duration and boundary conditions is shown in Figure \ref{bcandtime}.  

The worst case scenario leading to the smallest maximum safe input power density occurs when the cochlear walls are assumed to be adiabatic such that the insertion opening is the only boundary that enables heat transfer with the surroundings. At the other end of the spectrum,  a boundary condition of isothermal cochlear walls at 37$^{\circ}$C results in the largest maximum safe input power density.~Indeed, in this case, it is assumed that the thermoregulatory mechanisms adjust the temperature of the cochlear walls to 37$^{\circ}$C such that all boundaries act as heat sinks. Assuming the cochlear walls are isothermal, or when the cochlea is embedded in the temporal bone, the temperature difference between the cochlea walls and the magnet, as well as the total heat transfer surface area, increase. Thus, the maximum safe input power density increases in comparison to the adiabatic cochlea boundary assumption.

When accounting for the bone around the cochlea, the maximum safe input power density is between the two extreme cases of adiabatic and fixed-temperature cochlear walls. The maximum safe input power density is nearly independent of the bone boundary condition for heating duration less than 50 s. This is explained by the fact that for the adiabatic boundary condition, the temperature at the bone boundary does not appreciably exceed 37$^{\circ}$C owing to the low thermal conductivity of bone. At heating duration greater than 50 s, the maximum input power density for the isothermal bone boundary case gradually exceeds the power density for the adiabatic bone. At longer heating periods, heat has more time to diffuse through the bone,  which causes the bone boundary temperature to exceed 37$^{\circ}$C. Consequently, the maximum safe input power density is lower in comparison to the isothermal bone wall assumption.~The consideration of the temporal bone surrounding the cochlea provides the most realistic maximum safe input power density evaluation, but the adiabatic cochlea assumption provides the highest safety factor.

The numerical results confirm that the maximum safe input power density decreases by increasing the heating duration. As shown in Figure \ref{bcandtime} , the input power density is inversely proportional to heating duration for the case of an adiabatic cochlea, as predicted in Eq. (\ref{relations-ch6}). The total energy density transferred to the magnet, which is the product of the maximum safe input power density and heating duration, is lower for shorter heating duration. As a result, a strong pulsed heat source may debond and release the magnet without transferring a large amount of energy to the cochlea. For short heating duration, the heat removal rate from the magnet and its immediate adjacent tissues is not sufficiently large to remove significant heat from the magnet. As a result, the temperature of the magnet and its immediate adjacent tissues rises quickly, while the regions farther from the magnet do not immediately respond to the heat. Thus, for short heating periods, even though the safe input power density is larger in comparison to longer heating duration, the total energy transferred to the cochlea is lower. 

It was previously shown that an input power density of 1.8 $\rm{\frac{MW}{m^3}}$ for 114 s is required to melt 1 $\rm{mm^3}$ of paraffin bonding the magnet to the EA (under the assumption that the boundaries of the paraffin are adiabatic and no heat is wasted) \cite{uncoiledcochlea, cochleasecondpaper}. Here, the minimum safe input power density for 114 s of heating is determined to be 20 $\rm{\frac{MW}{m^3}}$, which is more than 11 times larger than the required input power density to melt the paraffin.~It can therefore be concluded that the magnet can be detached from the EA without causing thermal trauma if paraffin is the bonding material. Figure \ref{bcandtime} is a useful guide to determine whether or not the magnet detachment process is thermally safe by comparing the data to the required input power density to melt a specific adhesive. 

In the rest of the paper, the heating duration is fixed at 60 s.~Varying the heating duration does not change the main conclusions of the parametric analysis. In addition, only the adiabatic cochlear wall boundary condition is considered since it produces the highest safety factor in terms of the maximum safe input power density.

\begin{figure}[h]
\begin{center}
  \includegraphics[width=4 in]{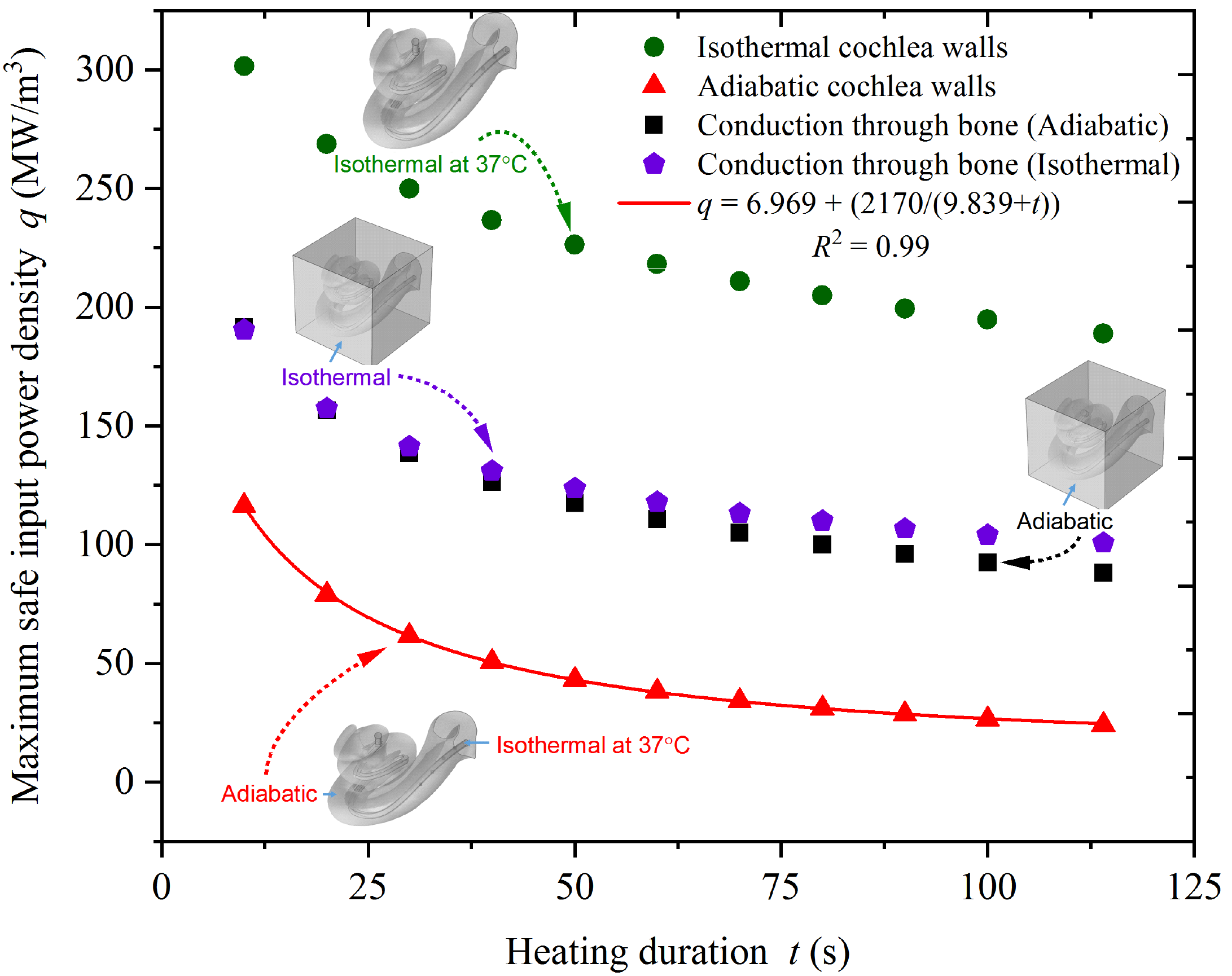}
  \caption{Maximum safe input power density as a function of the boundary condition and heating duration for 113 s $<$ CEM$_{43} <$ 114 s.}
  \label{bcandtime}
  \end{center}
\end{figure}

\subsection{Impact of electrode array (EA) radius and insertion depth}
\label{Impact of electrode array size: radius and insertion depth}
The embedded volume of the cochlear implant EA, which depends on its radius and insertion depth, is one of the parameters that impacts the maximum safe input power density. The effect of these characteristic lengths are discussed next.  

\subsubsection{Impact of EA radius}
\label{Impact of electrode array radius}
Cochlear implant EAs are fabricated as semi-conical objects because the diameter of the scala tympani decreases from the basal to the apical end.~The reported range of cochlear implant EA radius at the apical end of the cochlea varies from 0.125 mm to 0.25 mm \cite{DHANASINGH201793, implantdiameter}. In this paper, the cochlear implant EA is assumed to be cylindrical to reduce computational costs and complexities with a radius varying between 0.125 mm and 0.2 mm. The impact of EA radius on the maximum safe input power density is shown in Figure \ref{implantsize}. These numerical results are consistent with the dimensional analysis presented in Section \ref{Analytical and numerical parametric study-ch6}, which predicts that the maximum safe input power density is a quadratic function of EA radius. 

The implant EA is made of a semi-conical silicone structure, platinum wires, and electrodes. The finer details of the EA are not included in the geometrical model. The thermal effects of the wires and electrodes are taken into consideration by defining effective thermophysical properties for the EA. All of these properties are determined from the summation of mass fraction contributions for each of the individual component properties. The same method is applied to estimate the effective thermophysical properties of the perilymph with an embedded EA. The volumetric heat capacity is defined as the product of density ($\rho$) and specific heat ($c_p$).~For a fixed cochlea size, the total volume of the system of perilymph and EA is fixed. By increasing the EA radius, the volume of perilymph within the cochlea decreases. Heat is transferred from the hot magnet to the system surroundings by conduction through the EA and the perilymph. 

By increasing the radius of the implant from 0.125 mm to 0.2 mm, the effective volumetric heat capacity ($\rho c_p$) of the EA and perilymph system increases from $\sim$ 1.6 $\rm{\frac{MJ}{m^3 \cdot K}}$ to $\sim$ 21.3 $\rm{\frac{MJ}{m^3 \cdot K}}$, whereas the effective thermal conductivity $\kappa$ increases from 6.1 $\rm{\frac{W}{m \cdot K}}$ to 11.5 $\rm{\frac{W}{m \cdot K}}$. As a result, by increasing the radius of the implant EA, the heat removed from the magnet and the maximum safe input power density that can be applied to detach the magnet increase. 

\begin{figure}[h]
\begin{center}
  \includegraphics[width=4 in]{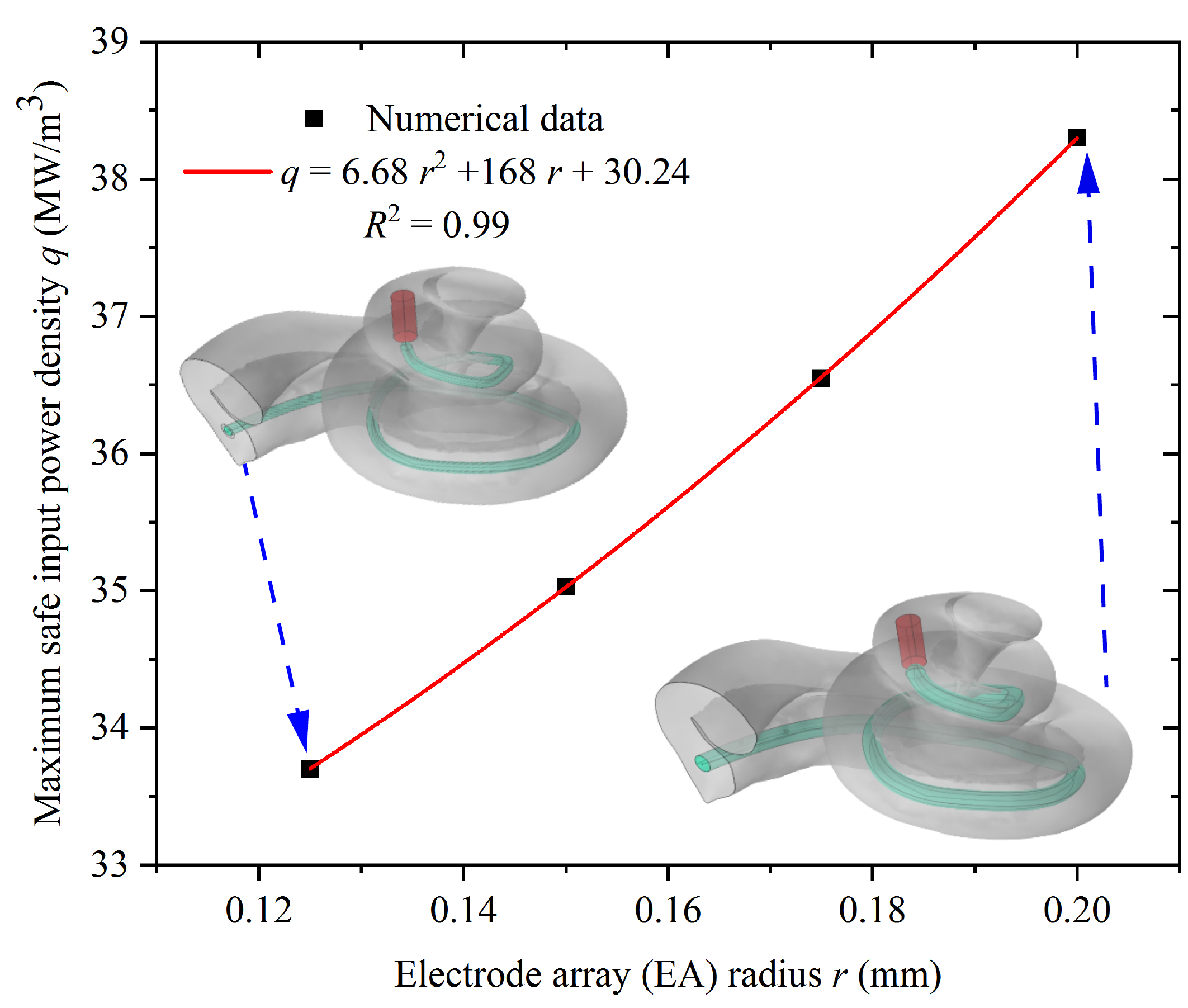}
  \caption{Maximum safe input power density as a function of the EA radius for a 60 s heating duration and an adiabatic boundary condition for the cochlear walls.}
  \label{implantsize}
  \end{center}
\end{figure}

\subsubsection{Impact of EA insertion depth}
\label{Depth of insertion}
The insertion depth of a cochlear implant EA, which influences the effective heat capacity of the cochlea, varies from $\sim$ 15 mm to $\sim$ 31.5 mm \cite{DHANASINGH201793}. At a fixed EA radius, increasing the insertion depth (i.e., the length of the EA inside the cochlea) from 10.53 mm to 22.5 mm, increases the effective volumetric heat capacity ($\rho c_p$) by 24.5$\%$. Even if the effective thermal conductivity increases as well, the divergence of the temperature gradient ($\nabla \cdot (\nabla T)$) between the magnet (heat source) and the isothermal insertion opening (heat sink) decreases by 64$\%$, which causes a drop in heat removal through the EA and, consequently, a drop of the maximum safe input power density. This result is in agreement with the dimensional analysis that shows that the power density is primarily inversely proportional to  EA insertion depth (see Figure \ref{insertiondepth}). The EA acts as a fin that removes heat from the cochlea, and, when the EA is shorter, it is more effective at heat removal because the temperature gradient is larger.~Alternatively, increasing the length of the implant EA inserted within the cochlea increases the distance between the magnet (heat source) and the isothermal insertion opening (heat sink), which increases the thermal resistance. 

\begin{figure}[h]
\begin{center}
  \includegraphics[width=4 in]{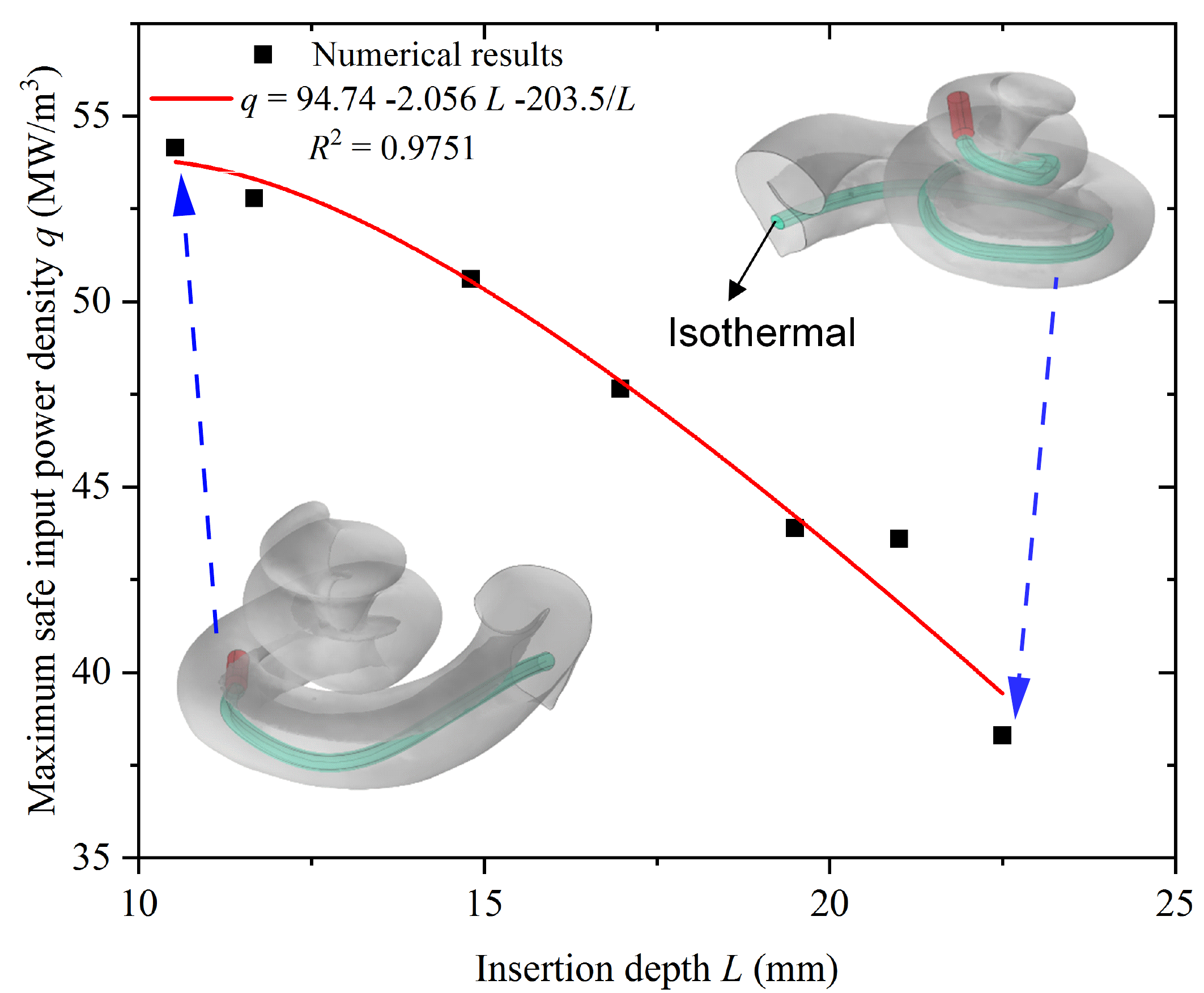}
  \caption{Maximum safe input power density as a function of the EA insertion depth for a 60 s heating duration and an adiabatic boundary condition at the cochlear walls.}
  \label{insertiondepth}
  \end{center}
\end{figure}

\subsection{Impact of cochlea size}
\label{Impact of cochlea size}
The size of the cochlea varies between individuals \cite{cochleasi}.~As such, it is important to address the impact of cochlea size on the maximum safe input power density to detach the magnet.~The impact of cochlea size is studied by fixing the radius and location of the EA and the size and location of the magnet at constant values and scaling the 3D model isometrically from $\sim$ 72.42 $\rm{mm^3}$ to $\sim$103.85 $\rm{mm^3}$ (see Figure \ref{cochleascale}). A larger maximum safe input power density may be applied for an individual with a larger cochlea, because the volume of the perilymph surrounding the magnet increases with increasing cochlea size. Perilymph is a heat sink that absorbs heat from the magnet. 
An increase in cochlea volume from 72.42 $\rm{mm^3}$ to 103.85 $\rm{mm^3}$ decreases the effective thermal conductivity by approximately 13.2$\%$ and increases the effective heat capacity ($\rho c_p V$) by approximately 27.7$\%$. Thus, more heat is absorbed by the perilymph inside the cochlea, and the competing effects of reduced thermal conductivity and increased heat capacity lead to a linear increase in the maximum safe input power density. The linear increase of maximum input power density with cochlea volume observed in Figure \ref{cochleascale} is consistent with the dimensional analysis of Section \ref{Analytical and numerical parametric study-ch6}.

\begin{figure}[h]
\begin{center}
  \includegraphics[width=4 in]{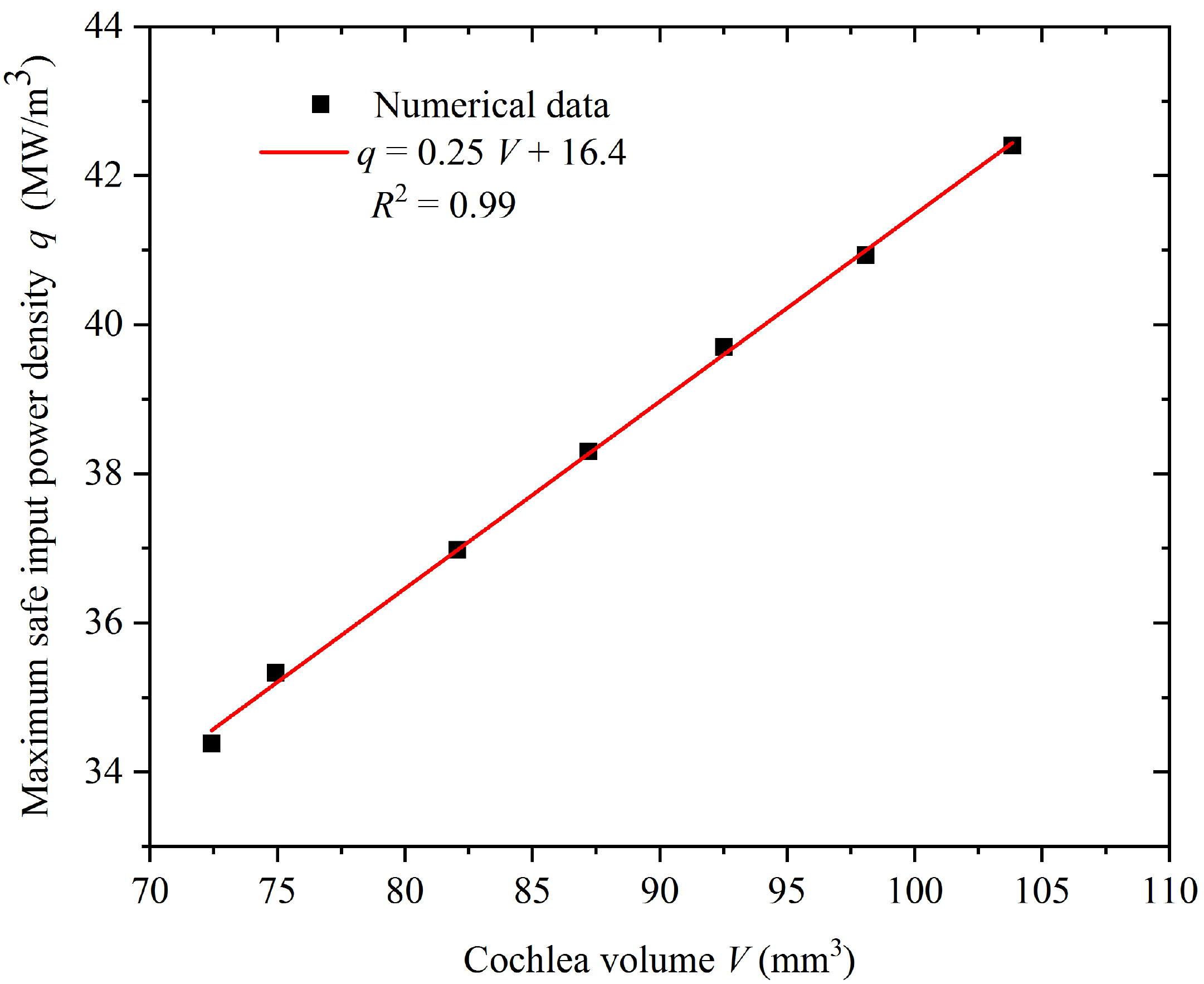}
  \caption{Maximum safe input power density as a function of the cochlea volume for fixed sizes of the EA and magnet. The results are for a 60 s heating duration and an adiabatic boundary condition at the cochlear walls.}
  \label{cochleascale}
  \end{center}
\end{figure}

\subsection{Impact of cochlear fluid}
\label{Impact of fluid}
The insertion of an EA forces some perilymph from the cochlea and may introduce air bubbles within the reamining fluid.~In addition, studies have shown that fluids such as glycerol decrease the insertion force required during cochlear implant surgery \cite{fluidimpact}. Thus, the cochlea may contain fluids ranging from perilymph to air and these fluids represent another parameter that can impact cochlea thermal performance. Even if glycerol decreases the insertion force \cite{fluidimpact}, it is not a good candidate from the thermal viewpoint due to its lower specific heat and thermal conductivity compared to perilymph (see Figure \ref{implantoffluid} and Table \ref{table}). A soap solution may also decrease the insertion force and its thermal properties are close to that of perilymph. Air is the worst fluid thermally because of its low specific heat, density, and thermal conductivity. Thus, during the magnet detachment, it is important to remove all air bubbles to avoid thermal trauma. 
\begin{figure}[h]
\begin{center}
  \includegraphics[width=4 in]{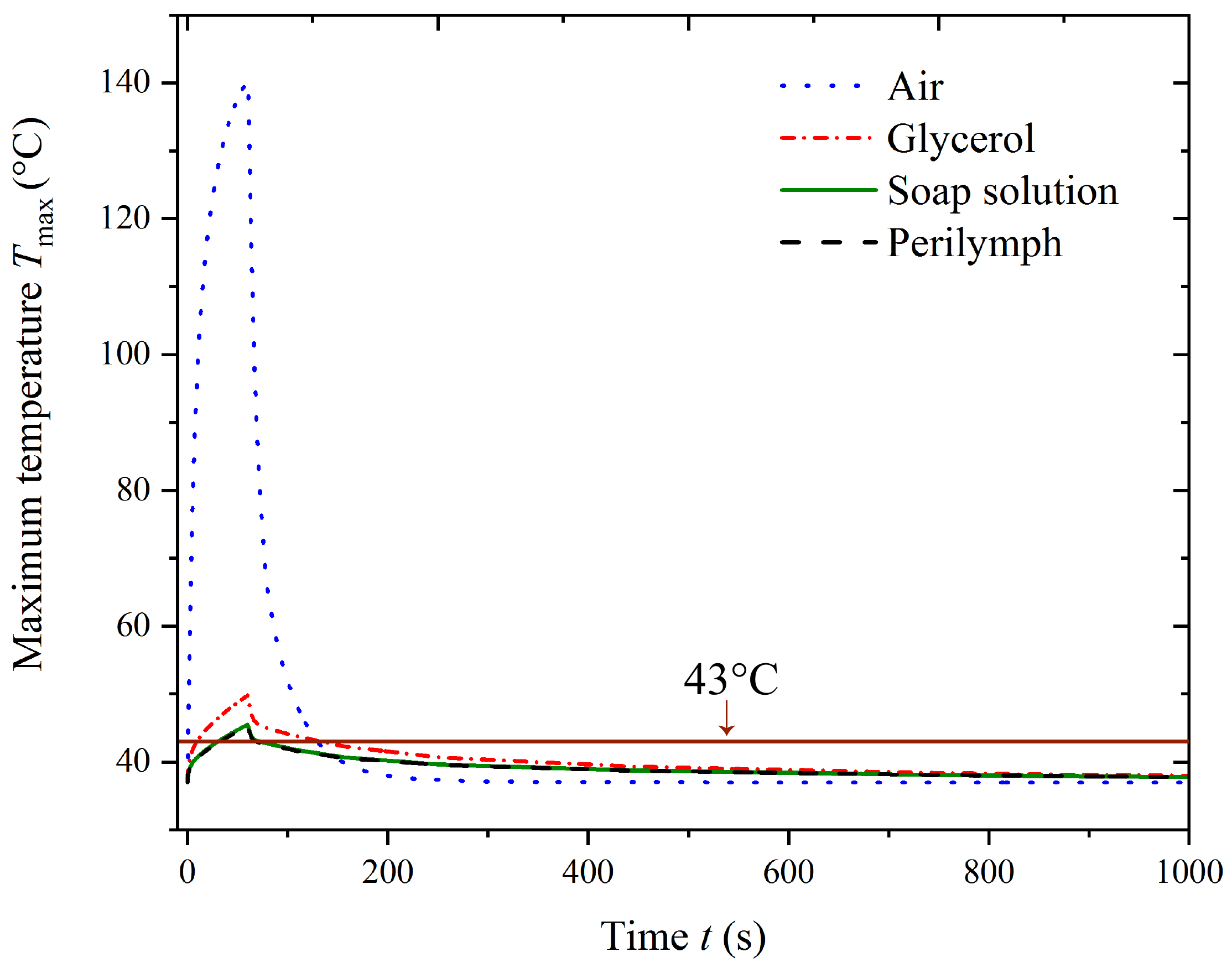}
  \caption{Maximum temperature within the cochlea as a function of the fluid for a 60 s heating duration, an adiabatic boundary condition at the cochlear walls, and an input power density $q$ = 38.3$\rm{\frac{W}{m^3}}$.}
  \label{implantoffluid}
  \end{center}
\end{figure}
\subsection{Impact of magnet size}
The size of the magnet driving the EA in cochlear implant surgeries may be adjusted based on the magnitude of the external electromagnetic field \cite{Bruns}.~The force applied on the magnet to steer it through cochlear turns is a function of the strength of the electromagnetic field and the surface area of the magnet. For instance, to provide a fixed force, a larger magnet with larger surface area requires a smaller magnetic field \cite{Bruns}, which mitigates overheating of the device that generates the electromagnetic field \cite{confPaper}. The maximum safe input power density is lower for a larger magnet (see Figures \ref{magnetsize} and \ref{magnetvolume-ch6}). Magnet size is described in Figure \ref{magnetsize} by its characteristic length, defined as the ratio of the magnet volume to its surface area $L_c=\frac{V_{\rm{magnet}}}{A_{\rm{magnet}}}$. In this paper, four magnet sizes are considered based on reported sizes used in magnetic insertion of cochlear implant EAs \cite{Bruns, Leon, Leon1, Leon2}. All magnets are assumed to be cylindrical. 

The magnet is located at the apical end of the cochlea. Increasing magnet size reduces the volume of perilymph surrounding the magnet. As a result, less heat is removed from the magnet by the perilymph and, consequently, the local temperature increases. Therefore, the rapid increase in local temperature limits the magnitude of the safe input power density. The numerical results are consistent with predictions from the dimensional analysis, which indicate that input power density decreases inversely with magnet volume (see Figure \ref{magnetvolume-ch6}).

\begin{figure}[h]
\begin{center}
  \includegraphics[width=4 in]{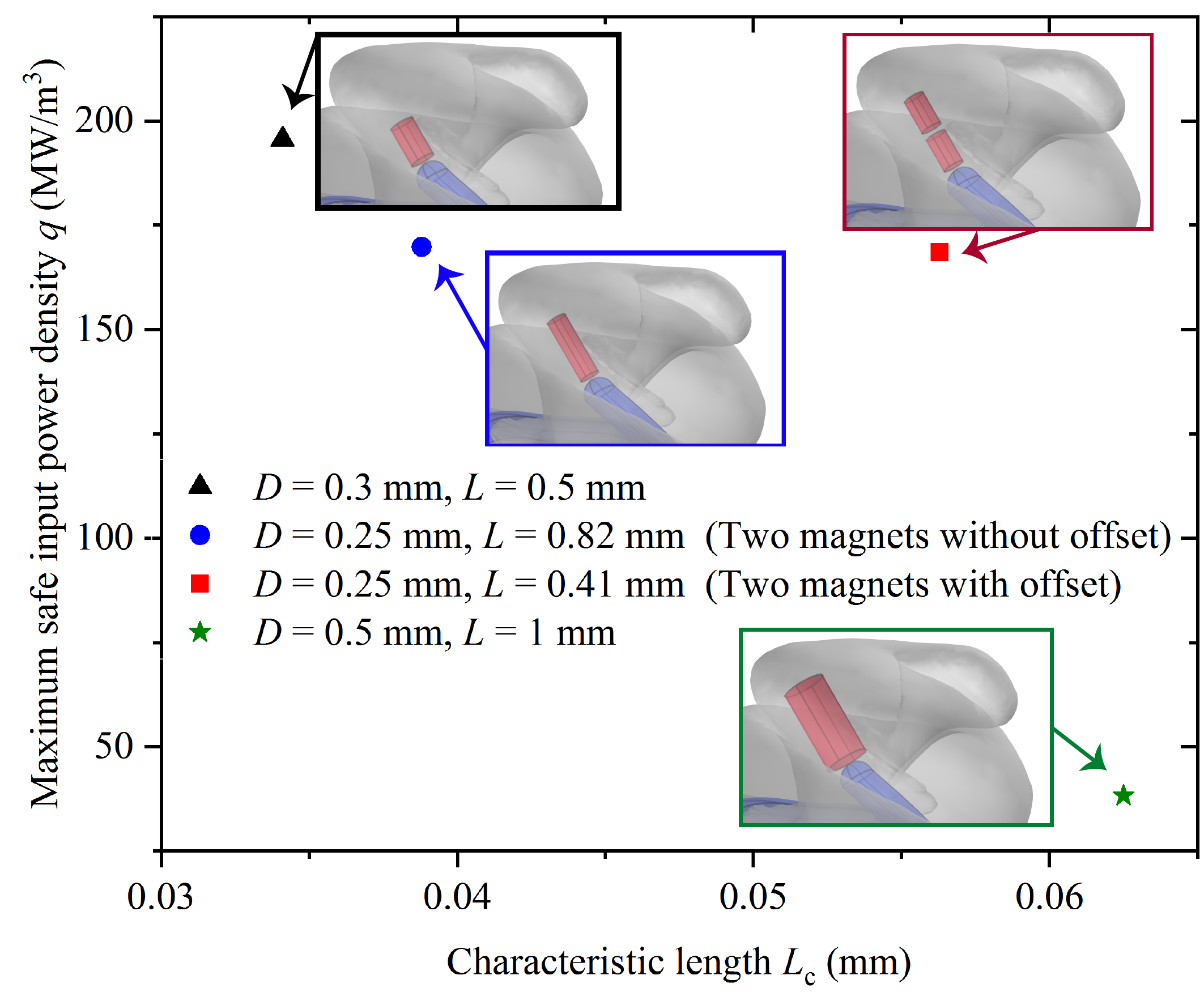}
  \caption{Maximum safe input power density as a function of the magnet characteristic length for a 60 s heating duration and an adiabatic boundary condition at the cochlear walls.}
  \label{magnetsize}
  \end{center}
\end{figure}

\begin{figure}[h]
\begin{center}
  \includegraphics[width=4 in]{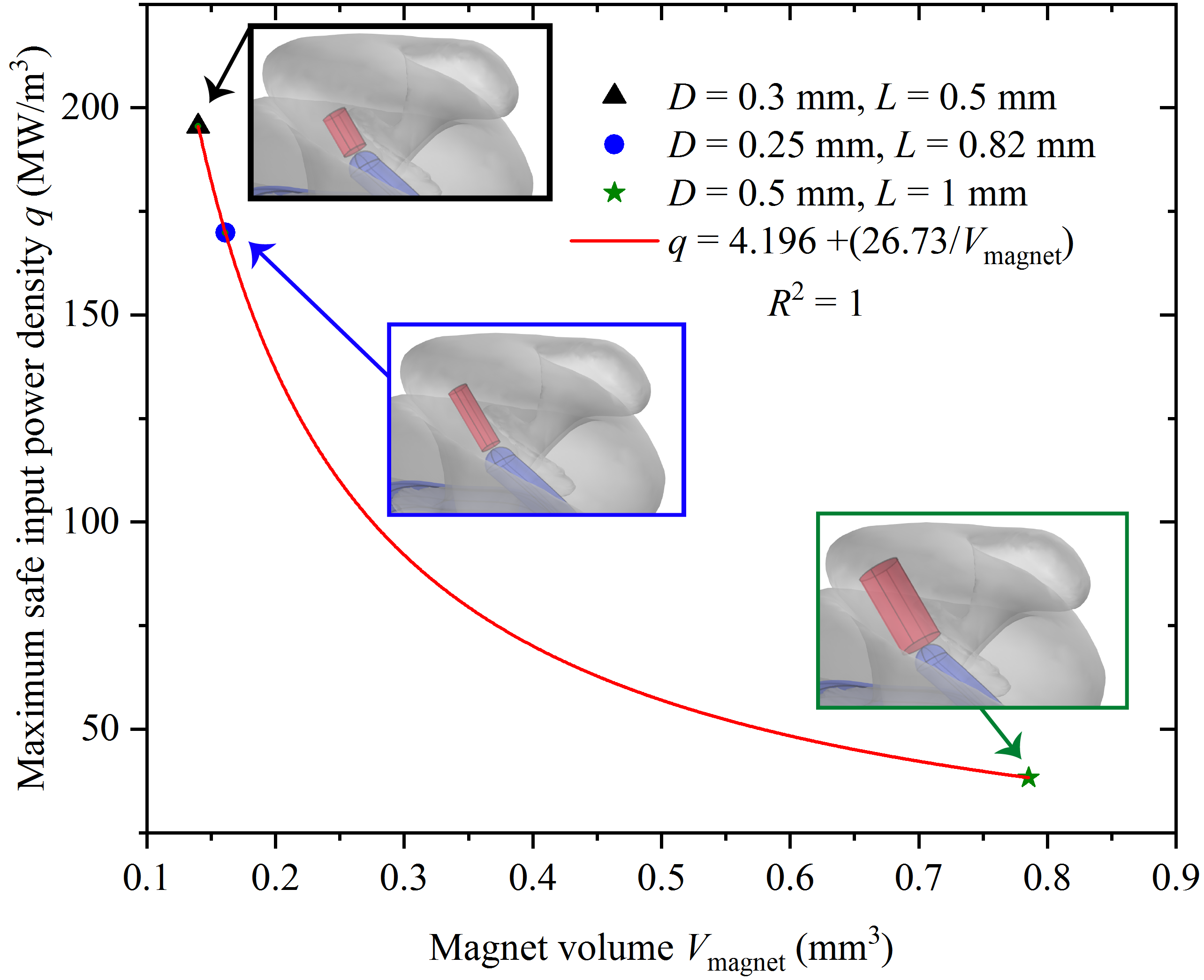}
  \caption{Maximum safe input power density as a function of the magnet volume for a 60 s heating duration and an adiabatic boundary condition at the cochlear walls.}
  \label{magnetvolume-ch6}
  \end{center}
\end{figure}

\section*{Conclusions}
A validated 3D heat transfer model of the cochlea has been applied to evaluate the impact of the cochlea thermal boundary condition, heating duration, the size and insertion depth of an implant EA, cochlea size, magnet size, and cochlear fluid on the maximum safe input power density to detach the magnet during magnetic cochlear implant surgery. The dimensional analysis confirmed the results of the computational study that maximum safe input power density is a quadratic function of the EA radius and a linear function of cochlea size, while it is inversely proportional to the insertion depth, magnet size, and heating duration.~As expected, the maximum safe input power density decreases with the heating period. The best fluid to use for the magnet detachment process is either perilymph or a soap solution. The safest input power density is calculated by assuming an adiabatic boundary condition for the cochlear walls. 

The design of the magnetic insertion process for cochlear implants requires a multidisciplinary approach. In addition to the thermal impact of the aforementioned parameters, the design requires knowledge of the electromagnetic field strength, cochlea dimensions, and desirable insertion depth, for example.~One solution is to consider the worst case scenario with the highest safety factor for all possible scenarios. For instance, the maximum safe input power density could be limited to its safest value and then it could be used as a standard to design the magnetic insertion procedure. For the safest thermal outcome, the adiabatic boundary condition for the cochlear walls, the smallest possible cochlea size, the smallest possible EA radius, the largest possible magnet, and the shortest heating duration should be applied to calculate the safest maximum input power density to detach and remove the magnet.

As future research, an optimization patient-specific algorithm that considers all the constraints and criteria involved in the thermal analysis, magnetic insertion procedure design, and medical requirements can be developed using the 3D heat transfer model presented in this paper.

\section*{Acknowledgment}
Research reported in this publication was supported by the National Institute on Deafness and Other Communication Disorders of the National Institutes of Health under Award Number R01DC013168. The content is solely the responsibility of the authors and does not necessarily represent the official views of the National Institutes of Health. We also acknowledge the Center for High Performance Computing at the University of Utah for their support and resources, and in particular thank Dr.\ Martin Cuma for his assistance in preparing the geometrical model of the cochlea.

\bibliography{cochlea3.bib}

\end{document}